\documentstyle[epsfig,preprint,aps]{revtex}

\def\bm{\boldmath}
\def\be{\begin{eqation}}
\def\ee{\end{eqation}}
\def\ben{\begin{eqnarray}}
\def\een{\end{eqnarray}}
\def\non{\nonumber}
\def\hs{\hspace{0.5em}}

\begin{document}
\tighten
\title{\bm Determination of the $\Theta^+$ parity   
from $\gamma n \to K^- K^+ n$}
\author{K. Nakayama,  K. Tsushima\footnote{From Dec. 1, 2003, 
Inst. de Fisica Teorica, UNESP, Rua Pamplona 145, 01405-900, 
Sao Paulo-SP, Brazil.}}
\address{$^1$Department of Physics and Astronomy, University of Georgia, 
Athens, Georgia 30602, USA }

\date{\today}
\maketitle


\begin{abstract}
It is demonstrated that measurements of photon asymmetry 
in the $\gamma  n \to K^-  K^+  n$ reaction, can most likely determine 
the parity of the newly discovered $\Theta^+$ pentaquark. We predict that 
if the parity of $\Theta^+$ is positive, the photon asymmetry is 
significantly positive; if the parity is negative, the photon asymmetry is 
significantly negative. If the background contribution is large,   
the photon asymmetry may become very small in magnitude, thereby making it 
difficult to distinguish between the positive and negative parity results. 
However, even in this case, a combined analysis of the $(K^+ n)$ invariant 
mass distribution and photon asymmetry should allow a  
determination of the parity of $\Theta^+$.
\\ \\
{\it PACS number(s)}: 13.60.Rj, 14.20.Gk, 14.20.-c, 13.88.+e
\end{abstract}
%
\newpage
The recent discovery of the pentaquark baryon $\Theta^+$ with the strangeness 
quantum number $S = +1$, mass $m_{\Theta^+}=1.54 \pm 0.01 GeV$ and width 
$\Gamma<25 MeV$ by the LEPS Collaboration at SPring-8 ~\cite{LEPS}
has triggered an intensive investigation of exotic baryons.
The $\Theta^+$ baryon (renamed from $Z^+$) has been also identified by 
other experimental groups~\cite{DIANA,CLAS,SAPHIR,neutrino}.
Furthermore, the NA49 Collaboration~\cite{NA49} has reported the finding 
of another pentaquark baryon, $\Xi_{3/2}$, with $S = -2$. The pentaquark 
$\Theta^+ (Z^+)$ was predicted by Diakonov, Petrov and Polyakov~\cite{Diakonov} 
in 1997 in the chiral soliton model as the lowest member of an anti-decuplet 
baryons. The existence of such exotic baryons was discussed even earlier 
by a number of authors~\cite{Golowich,Strottman,Chemtob}. Before the 
announcement of the LEPS Collaboration's discovery, a theoretical study 
of the $\Theta^+ (Z^+)$ was also made based on the Skyrme model~\cite{Weigel}. 
Also, investigation about some experimental possibilities have been made in 
Ref.~\cite{Alex}.

Although the existence of $\Theta^+$ has been confirmed experimentally, 
many of its basic properties such as its quantum numbers 
remain undetermined. From the lack of a signal in the $(K^+p)$ 
invariant mass distribution in $\gamma p \to K^- K^+ p$, the SAPHIR 
Collaboration~\cite{SAPHIR} has concluded 
that $\Theta^+$ should be an isoscalar state. Currently available data do not 
allow for the determination of either its spin or its parity. Consequently, 
many theoretical studies of $\Theta^+$, based on a number of different 
approaches, are available aimed at establishing these properties
\cite{Jaffe,Karliner,group,Riska,Carlson,Cheung,Kim,QCDsum1,QCDsum2,QCDsum3,lattice,Skyrme,tube,Capstic,Cohen,Bijker}.
The results, however, are largely controversial. Naive SU(6) quark 
model as well as QCD sum rule~\cite{QCDsum1,QCDsum3} 
calculations predict a spin 1/2 negative parity state. 
Also, recent 
quenched lattice QCD calculations~\cite{lattice} identified the 
spin 1/2 $\Theta^+$ as the lowest mass ($1539 \pm 50$ MeV)  
isoscalar negative parity state; a state with either 
isospin 1 and/or positive parity lies at a higher mass. 
In contrast, chiral/Skyrme soliton 
models~\cite{Diakonov,Weigel} and correlated quark 
models~\cite{Jaffe,group} predict a spin 1/2 positive parity isoscalar state. 
Goldstone boson exchange \cite{Riska,Kim} and color magnetic exchange quark 
models~\cite{Kim} also predict a positive parity for $\Theta^+$. 
Yet, in another work~\cite{Capstic}, the observed narrow 
width of $\Theta^+$ has been ascribed 
to this baryon possibly belonging to an isotensor multiplet. 
Concerning the structure of $\Theta^+$, an interesting possibility of a 
diquark-diquark-antiquark ($[ud][ud]\bar{s}$) 
flavor structure for $\Theta^+$ has been 
introduced in Ref.\cite{Jaffe} 
(see also Refs.\cite{Cheung,QCDsum2}). On the other hand, in Ref.\cite{Karliner}
it is interpreted as having a diquark-triquark ($[ud][ud\bar{s}]$) structure. 
In addition to these theoretical efforts, there exists other theoretical 
studies 
addressing the reaction aspects involving 
$\Theta^+$\cite{KN,Polyakov,Ko1,RHIC,Ko2,Hydo,Nam,Oh}.
In particular, Refs.~\cite{Hydo,Nam,Oh} explore the possibility 
of determining its quantum numbers (parity) experimentally. 
However, none of these calculations have 
considered the three/four-body final states which are involved in the actual 
experiments; the $\Theta^+$ baryon is present only in the 
intermediate state in these experiments.   

In the present study we focus on the $\gamma n \to K^- K^+ n$ reaction, 
which has been investigated experimentally by 
the LEPS Collaboration~\cite{LEPS}, and demonstrate that measurements 
of the photon asymmetry in conjunction with 
$(K^+n)$ invariant mass distribution can most likely determine 
the parity of an isospin $0$ and spin $1/2$ $\Theta^+$ pentaquark. 

\begin{figure}
\begin{center}
\vspace*{-2cm}
\epsfig{file=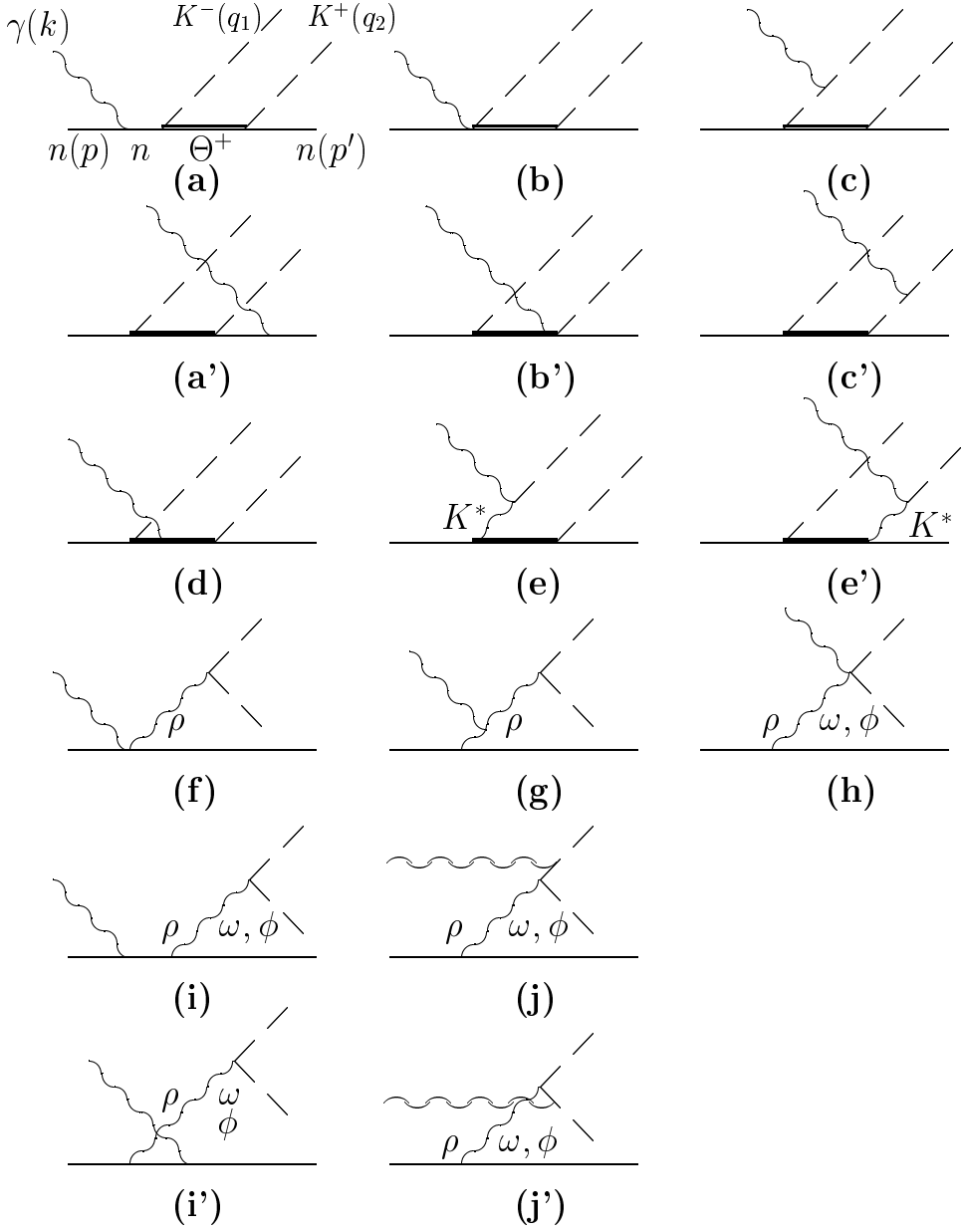,width=17cm}
\vspace{-3cm}
\caption{Processes considered in this study: 
$K$ [(a)-(d) and (a')-(c')] and $K^*$ [(e) and (e')] contributions, 
and the background due to 
the $\rho$ [(f)-(j), (i') and (j')], 
$\omega$ and $\phi$ [(h)-(j), (i') and (j')] exchanges, and the 
$\Sigma(1197)^-$ and $\Sigma(1660)^-$ intermediate states 
[by the replacements, $\Theta^+ \to \Sigma(1197)^-$ or $\Sigma(1660)^-, 
K^- \leftrightarrow K^+$ in (a)-(e) and (a')-(e')]. 
}
\label{diagrams}
\end{center}
\end{figure}

In Fig.\ref{diagrams} we depict the processes considered in the present work. 
In order to investigate the effect of various reaction mechanisms, we group the 
diagrams (a)-(d) and (a')-(c') together and refer to them as the $K$ contribution. 
Diagrams (e)
and (e') are referred to as the $K^*$ contribution. Together ($K+K^*$), 
they constitute the $\Theta^+$ (resonance) contribution. Although experimental 
evidence~\cite{LEPS,DIANA,CLAS,SAPHIR,neutrino} suggests a strong 
$NK\Theta^+$ coupling, the $NK^*\Theta^+$ coupling may also be important in the
excitation of the $\Theta^+$~\cite{Oh}. In order to obtain results which can be 
compared directly to those measured, the background contribution 
needs to be included in the calculation. Presently, however, there is a large
uncertainty in the background contribution.
We therefore make a rough estimate of 
its effect relevant for the present study. We consider the $\rho$, 
$\omega$  and $\phi$ meson exchange diagrams (Figs.\ref{diagrams}(f)-(j), (i'), 
and (j')). In addition, we also include the $\Sigma(1197)^-$ and $\Sigma(1660)^-$
contributions for the background. They are obtained from the diagrams 
Figs.\ref{diagrams}(a)-(e) and (a')-(e') by replacing $\Theta^+$ by $\Sigma(1197)^-$, 
$\Sigma(1660)^-$ and interchanging $K^-(q_1)$ with $K^+(q_2)$. 
Although the decay channels $\rho \to \bar K K$ and $\omega \to \bar K K$ are 
kinematically closed, these vector meson exchanges 
contribute largely to the background due to their strong coupling to nucleons.
The $\Sigma(1197)^-$ is expected to have a strong coupling to the $N\bar{K}$ 
channel according to the hyperon-nucleon ($YN$) interaction models~\cite{YNmodel}, 
and the $\Sigma(1660)^-$ has an appreciable decay branch to the $N\bar{K}$ 
channel~\cite{PDG}. It is possible to remove the $\phi$ exchange contribution
from the experimental background by measuring the $(K^+K^-)$ invariant mass 
distribution and rejecting the events corresponding to it ($\phi$) as has been done in 
Refs.~\cite{LEPS,CLAS,SAPHIR,Nakano}. In principle, the $\Sigma(1660)^-$ 
contribution may also be removed from the experimental background by measuring the
$(K^-n)$ invariant mass distribution. However, this may not be practical due to the
relatively large width of this resonance.

We work with an effective Lagrangian at the hadronic level.
The hadronic parts of the interaction Lagrangian are given by: 
\ben
{\cal L}^\pm_{NKR} &=& -g_{NKR} 
\bar{N} \left[ i\lambda\Gamma^\pm R  
\pm \frac{1-\lambda}{m_R \pm m_N}\Gamma^\pm_\mu R \partial^\mu 
\right] K + h.c.,
\non
\\
& &\hspace{17em}
[R=\Theta^+,\vec\tau\cdot\vec\Sigma(1197),\vec\tau\cdot\vec\Sigma(1660)], 
\label{NKTheta}
\\
{\cal L}^\pm_{NK^*R} &=& \frac{g_{NK^*R}}{(m_R+m_N)^2} 
\non\\
&\times&\bar{N}\left[ \Gamma_\mu^\mp R\partial^2 
\mp (m_R \mp m_N) i\Gamma^\mp R\partial_\mu  
+ (m_R + m_N)\kappa^*\Gamma^\mp\sigma_{\mu\nu}R\partial^\nu
\right]K^{*\mu} + h.c., 
\non
\\
& &\hspace{17em}
[R=\Theta^+,\vec\tau\cdot\vec\Sigma(1197),\vec\tau\cdot\vec\Sigma(1660)],
\label{NK*Theta}
\\
{\rm with}& &  
\Gamma^\pm = {\gamma_5 \choose 1}, 
\Gamma_\mu^\pm = {\gamma_5\gamma_\mu \choose \gamma_\mu}, \non
\\
{\cal L}_{VNN} &=& -g_{VNN} 
\bar{N}\left\{ \left[ \gamma_\mu 
- \frac{\kappa_VNN}{2m_N}\sigma_{\mu\nu}\partial^\nu\right] V^\mu \right\} N, 
\hs 
[V^\mu = (\vec{\tau}\cdot\vec{\rho}^{\,\mu}),\omega^\mu,\phi^\mu], 
\label{VNN}
\\
{\cal L}_{VKK} &=& -ig_{VKK}
\left[ \bar{K}V^\mu(\partial^\mu K) 
- (\partial_\mu \bar{K})V^\mu K \right], 
\hs
[V^\mu = (\vec{\tau}\cdot\vec{\rho}^{\,\mu}),\omega^\mu,\phi^\mu],
\label{VKK}
\een
where $K^T = (K^+, K^0)$ and $\bar{K} = (K^-, \bar{K}^0)$ and similarly for 
$(K^{*\mu})^T$ and $\bar{K}^{*\mu}$, with $T$ the transpose operation.
The superscript $\pm$ in Eqs.(\ref{NKTheta},\ref{NK*Theta}) as well as 
in Eq.(\ref{NKRgamma}) below stands for the 
positive ($+$) and negative ($-$) parity baryon $R$. 
The parameter $\lambda$ appearing in Eq.(\ref{NKTheta}) controls 
the pseudoscalar-pseudovector (ps-pv) [scalar-vector] admixture 
of the $KNR$ coupling for the positive [negative] parity baryon $R$.

In addition, the electromagnetic 
parts of the interaction Lagrangian are given by: 
\ben
{\cal L}_{BB\gamma} &=& 
- \bar{B} e_B \left\{ \left[ \gamma_\mu
- \frac{\kappa_B}{2m_N}\sigma_{\mu\nu}\partial^\nu\right] 
A^\mu \right\} B,\hs [B = \Theta^+,N,\Sigma(1197)^-,\Sigma(1660)^-],
\label{BBgamma}
\\
{\cal L}_{KK\gamma} &=& -ie 
\left[ K^-(\partial_\mu K^+) - (\partial_\mu K^-)K^+ \right] A^\mu,
\label{KKgamma}
\\
{\cal L}_{NKR\gamma}^{\pm} &=& \mp ieg_{NKR} 
\left( \frac{1 - \lambda}{m_R \pm m_N} \right)
\bar{R} \Gamma^\pm_\mu N K^- A^\mu + h.c.,
\hs [R = \Theta^+,\Sigma(1197)^-,\Sigma(1660)^-], 
\label{NKRgamma}
\\
{\cal L}_{KK^*\gamma} &=& \left( \frac{g_{KK^*\gamma}}{m_K} \right)
\varepsilon^{\mu\nu\lambda\sigma} (\partial_\mu A_\nu) 
\left[ (\partial_\lambda K^-)K^{*+}_\sigma 
+ (\partial_\lambda \bar{K}^0)K^{*0}_\sigma \right] + h.c., 
\label{KK*gamma}
\\
{\cal L}_{\rho\rho\gamma} &=& -e 
\left\{ A^\mu [\vec{\rho}^{\,\nu}\times
(\partial_\mu\vec{\rho}_\nu - \partial_\nu\vec{\rho}_\mu)]_3
- (\partial^\mu A^\nu)[\vec{\rho}_\mu\times\vec{\rho}_\nu]_3 \right\}, 
\label{rhorhogamma}
\\
{\cal L}_{\rho NN\gamma} &=& e \frac{f_{\rho NN}}{2m_N}
\bar{N} \sigma_{\mu\nu} A^\mu (\vec{\rho}^{\,\nu}\times\vec{\tau})_3 N
\label{rhoNNgamma}
\\
{\cal L}_{VKK\gamma} &=& e g_{VKK}
K^- K^+ V^\mu A_\mu + h.c.,\hs [V^\mu=\rho^\mu_3,\omega^\mu,\phi^\mu],
\label{VKKgamma}
\een
where $e_B$ in Eq.(\ref{BBgamma}) is the electric charge operator of the 
baryon $B$ and, $e$, the proton charge. $A^\mu$ stands for the photon field.
Note that the same parameter $\lambda$ in Eq.(\ref{NKTheta}) also appears 
in Eq.(\ref{NKRgamma}). This is needed to ensure gauge invariance of the 
resulting reaction amplitude. The same is true for the coupling constant
$f_{\rho NN}\equiv g_{\rho NN}\kappa_\rho$ in Eqs.(\ref{VNN},\ref{rhoNNgamma}).

The values of the coupling constants in the above interaction Lagrangians  
are summarized in Table~\ref{cconst}. We utilize the relevant sources 
whenever available as indicated in the Table to determine these 
couplings. Most of those couplings that cannot be extracted from other 
sources, are estimated following Ref.~\cite{Nak1} from a systematic 
analysis based on SU(3) symmetry in conjunction with Okubo-Zweik-Iizuka (OZI) 
rule constraints~\cite{OZI} and some experimental data. The remaining few 
parameter values are simply assumed. The coupling constant 
$g_{NK\Theta}$ in Eq.(\ref{NKTheta}) is estimated using the upper limit of 
the decay width of $\Gamma_{(\Theta^+ \to K^+ n)} = 25 MeV$~\cite{LEPS}. The 
mixing parameter $\lambda$ in Eq.(\ref{NKTheta}) is treated as a free parameter; 
we consider both the extreme values of $\lambda=1$  and $\lambda=0$. Nothing 
is known about the coupling constant $g_{NK^*\Theta}$ in Eq.(\ref{NK*Theta}). 
Following  Ref.~\cite{Oh} we employ $g_{NK^*\Theta} \sim (1/2)g_{NK\Theta}$, 
assuming the same ratios obtained for the $NKY$ to $NK^*Y$ coupling constants 
($Y=\Sigma,\Lambda$) empirically~\cite{Workman}. We also consider 
$g_{NK^*\Theta} \sim - g_{NK\Theta}$, an estimate resulting from assuming the
same ratio obtained for the $NK\Sigma$ to $NK^*\Sigma$ coupling constants 
from SU(3) symmetry considerations (see Table~\ref{cconst}). Furthermore, 
both choices of the sign ($\pm$) for $g_{NK^*\Theta}$ are also considered. 
The tensor to vector coupling constant ratio $\kappa^*$ in Eq.(\ref{NK*Theta})
is treated as a free parameter; we consider $\kappa^* = -3, 0, +3$. In 
Table~\ref{cconst}, for the $D$ to $F$ admixture parameter 
$\beta = (D/F) / [1+(D/F)]$ in the
SU(3) baryon-baryon-pseudoscalar meson Lagrangian \cite{Nak1}, 
we use the value of $\beta \simeq 0.63$ \cite{Nak2} as obtained from 
averaging the values extracted from a systematic analysis of semileptonic 
hyperon decays \cite{Rat}. In the baryon-baryon-vector meson Lagrangian we 
take $\beta=0$ \cite{Nak1} which is a consequence of requiring spin 
independence for $BB\omega$ and $BB\phi$ couplings within the identically 
flavored baryons ($B=\Lambda,\Sigma$). For further details about the 
determination of the coupling constants, we refer to Ref.\cite{Nak1}.

It should be stressed that the present calculation cannot provide a 
quantitative prediction of the absolute value of the cross section. 
In phenomenological approaches like the present one, one usually 
introduces form factors at the hadronic vertices to
account for the composite nature of hadrons. It happens that little is 
known about the form factors needed in the present study, 
and the introduction of such form factors would significantly increase 
the number of unknown parameters in the model. Moreover, the
presence of form factors usually leads to the breaking of gauge invariance 
of the resulting amplitude and 
a proper restoration of gauge invariance is not a trivial 
task. Therefore, in the present study, we simply leave out the form 
factors. 
Accordingly, we expect that the present calculation should be more reliable
for relative quantities than for absolute cross sections. 
(In lowest order, the role of form factors would cancel exactly in 
calculations of relative quantities such as the photon asymmetry if the 
process were dominated by a single production mechanism. Incidentally, 
for the ($K^+n$) invariant mass around the resonance peak, diagram 
Fig.~\ref{diagrams}(b) 
dominates in the case of a positive parity $\Theta^+$; moreover, this 
diagram would not be affected much by the (off-shell) form factors, for 
$\Theta^+$ will be nearly on-shell. As we shall show, for a negative parity 
$\Theta^+$, the background contribution may be relatively large, in 
which case the influence of the form factors may be stronger.) 
A quantitative assessment of this reliability would require exploratory 
calculations which include a range of (unknown) form factors, in addition 
to properly preserving gauge invariance of the resulting amplitude. This 
is beyond the scope of this letter.

The primary focus of this study is on the photon asymmetry. 
It is defined as follows: 
let the 4-momenta of the photon and $K^-$ meson be  
$k^\mu \equiv (|\vec{k}|,0,0,|\vec{k}|)$ and 
$q^\mu_1 \equiv (q^0_1,q_{1x},0,q_{1z}) = (q^0_1,\vec{q}_1)$, 
respectively. Take the $y$-axis parallel to 
$(\vec{k} \times \vec{q}_1)$ and define the photon polarization vectors,
$\epsilon^\mu(\lambda_\gamma=+1) \equiv (0,0,1,0)$ and 
$\epsilon^\mu(\lambda_\gamma=-1) \equiv (0,1,0,0)$. 
Then, the photon asymmetry, $\Sigma$, is given by,
\ben
\Sigma &\equiv& 
\frac{d\sigma(\lambda_\gamma=+1) - d\sigma(\lambda_\gamma=-1)} 
{d\sigma(\lambda_\gamma=+1) + d\sigma(\lambda_\gamma=-1)}, 
\label{asymmetry}
\een
where $d\sigma(\lambda_\gamma) 
\equiv d^2\sigma(\lambda_\gamma)/[dm_{(K^+ n)} d\Omega_{K^-}]$ with 
$m_{(K^+n)}$ being the invariant mass of the $(K^+n)$ system.

\begin{figure}
\begin{center}
\vspace*{-1cm}
\epsfig{file=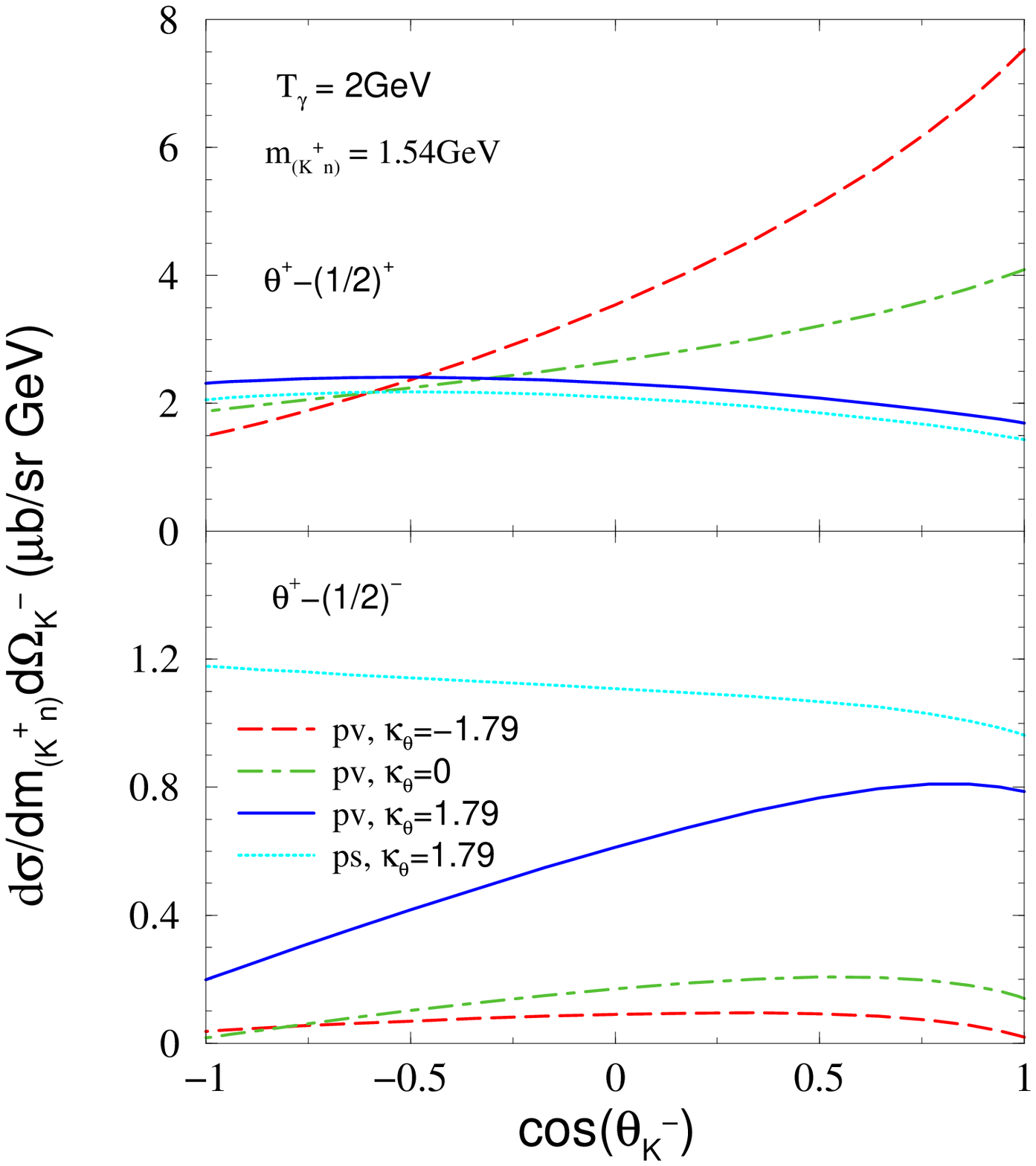,width=9.5cm,height=13cm,angle=-90}\\
\hspace*{-0.5cm}
\epsfig{file=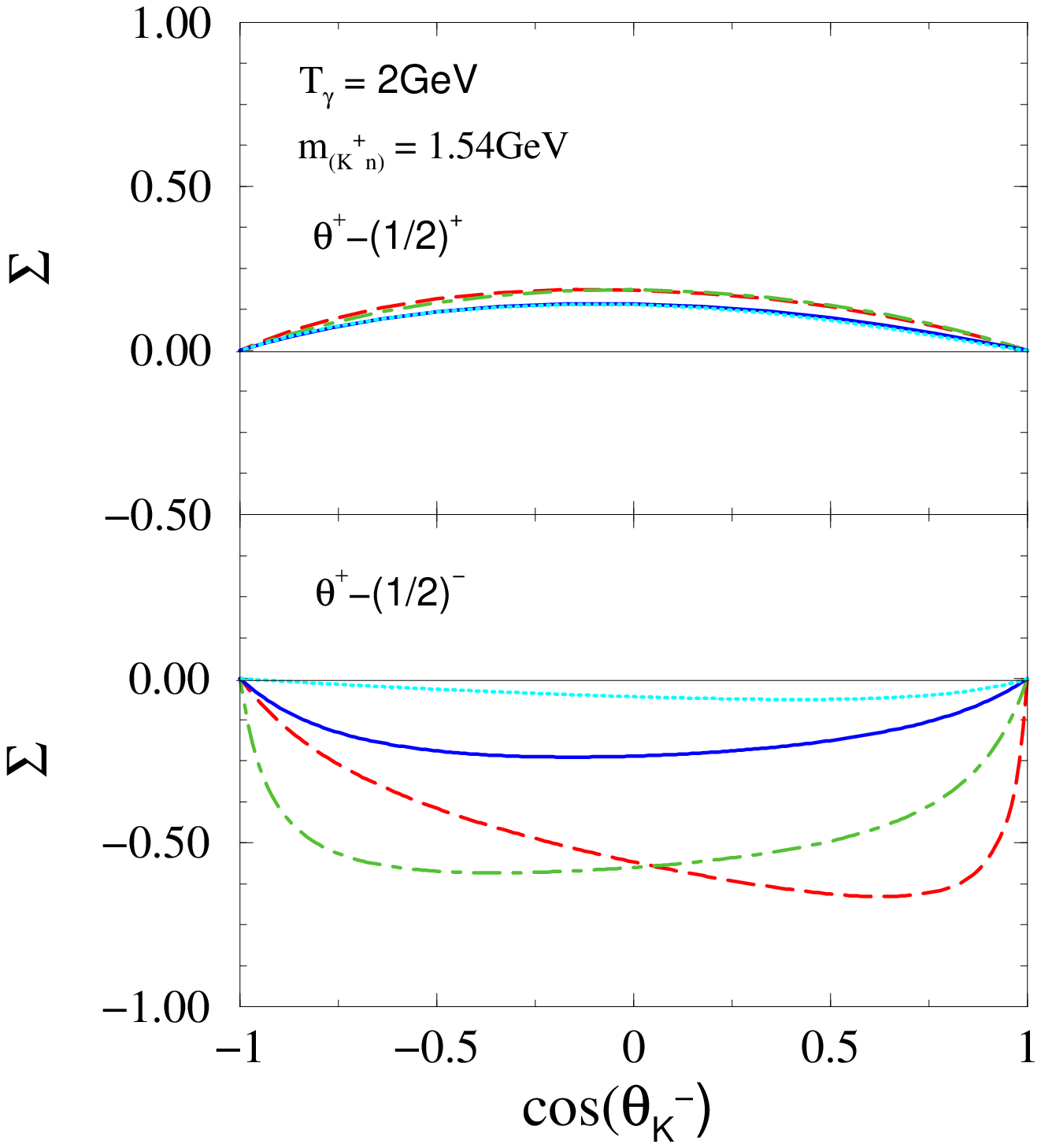,width=13.5cm,height=9.5cm}
\caption{$K^-$ angular distribution for double differential cross section 
(upper figure) and photon asymmetry $\Sigma$ (lower figure) in the 
$(\gamma n)$ center-of-mass frame calculated with the $K$ contribution 
alone as defined in the text, for both positive and negative parity cases of $\Theta^+$. 
The photon laboratory energy is $T_\gamma = 2$ GeV and the $(K^+n)$ invariant mass is
$m_{(K^+n)} = 1.54$ GeV. Different curves correspond to different choices of the 
parameter values $\lambda$ and $\kappa_{\Theta}$ in Eqs.~(\protect\ref{NKTheta}) and
(\protect\ref{BBgamma}), respectively.
}
\label{K_only}
\end{center}
\end{figure}

We now focus on the results of the present calculation.
Let's first concentrate on the $K$ contribution. 
Fig.\ref{K_only} shows the results for the (double differential) 
cross sections (upper figure) 
and photon asymmetries (lower figure) as a function 
of $K^-$ emission angle, $\cos(\theta_{K^-})$, in the overall 
center-of-mass (c.m.) frame at an incident photon laboratory  
energy of $T_\gamma = 2$ GeV and fixed $(K^+n)$ invariant mass  
$m_{(K^+n)} = 1.54$ GeV. Results are shown for both the positive 
($J^P=1/2^+$) and negative ($J^P=1/2^-$) parity choices of 
$\Theta^+$. Different curves correspond to different
values of the ps-pv (scalar-vector) mixing parameter $\lambda$ in 
Eq.(\ref{NKTheta}) and the anomalous magnetic moment  
$\kappa_\Theta$ of $\Theta^+$ in Eq.(\ref{BBgamma}). 
These are the only free parameters in the $K$ contribution. 
As can be seen, for the  cross section, 
the case of positive parity $\Theta^+$ (upper panel)
enhances the angular distribution at forward angles as the anomalous
magnetic moment $\kappa_\Theta$ decreases.
For the negative parity case (lower panel), however, the
effect of $\kappa_\Theta$ is just the opposite, i.e., the angular 
distribution is reduced at forward angles as $\kappa_\Theta$ decreases. 
A comparison of the solid ($\lambda=0$)
and dotted ($\lambda=1$) curves reveals the sensitivity of 
the angular distribution to the ps-pv (scalar-vector) 
mixing parameter $\lambda$. 
For the positive parity case, it is quite 
insensitive to this parameter, while for the negative parity case, 
the cross section is enhanced primarily at backward angles. 
Apart from the fact that the cross section is much smaller  
for the negative parity case of $\Theta^+$ than for the positive 
parity case - a feature that has been also pointed out 
in Ref.~\cite{Oh} - one can conclude that it will be difficult to 
determine the parity of $\Theta^+$ from the shape of the 
angular distribution. However, the situation is quite different  
for the photon asymmetry $\Sigma$. 
The lower figure in Fig.\ref{K_only} illustrates the sensitivity of 
the photon asymmetry $\Sigma$ to the only free parameters, 
$\lambda$ and $\kappa_\Theta$, of the calculation. One can see that for 
the positive parity case of $\Theta^+$ the photon asymmetry $\Sigma$ 
is always positive, while it is always negative for the negative 
parity case. Therefore, in contrast to the cross section, 
measurements of the photon asymmetry $\Sigma$ can potentially determine 
the parity of $\Theta^+$. Of course, other reaction mechanism(s) 
should be investigated before a more definitive statement can be made, and 
we now examine this.

\begin{figure}
\begin{center}
\vspace*{-1cm}
\epsfig{file=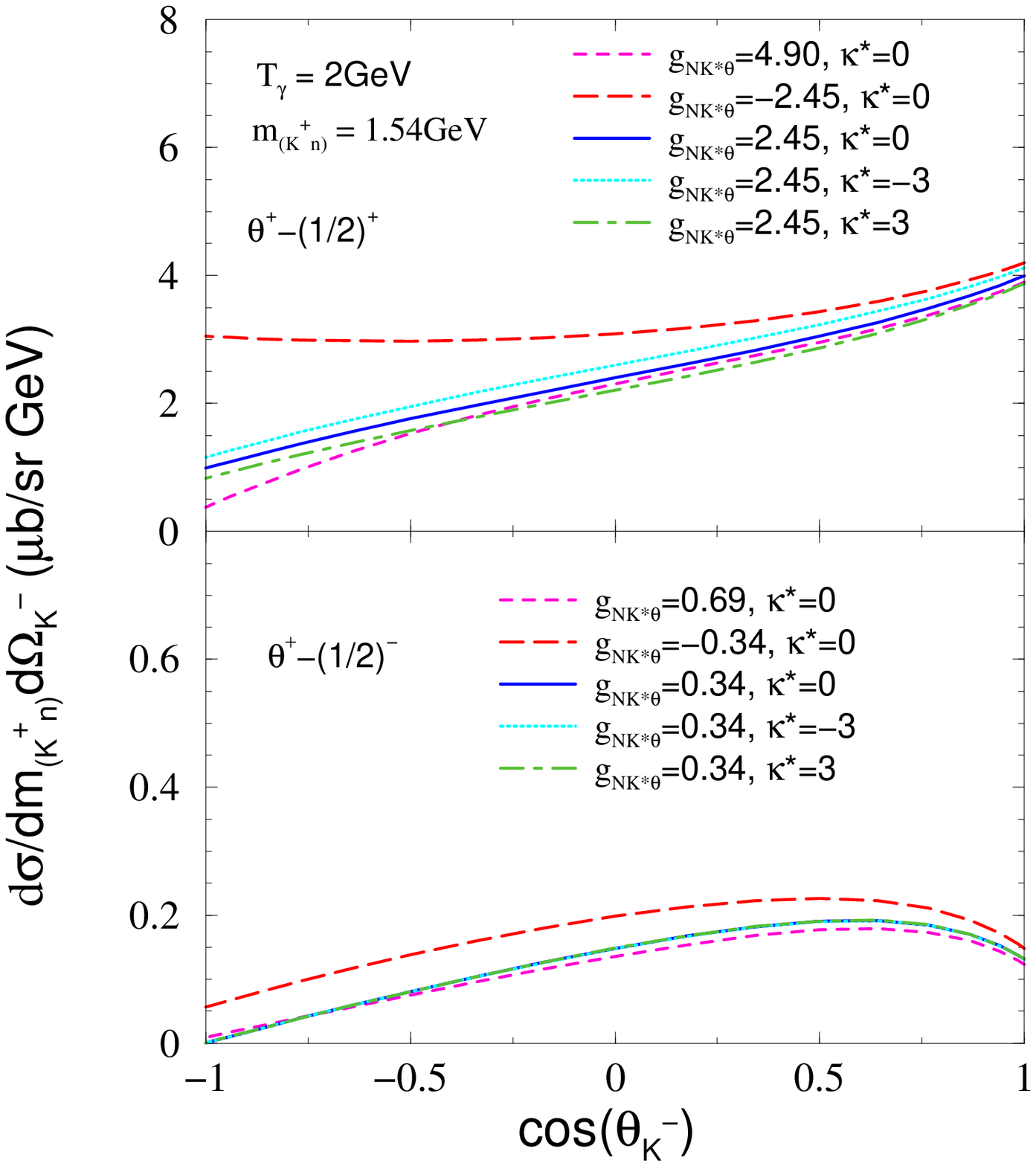,width=9.5cm,height=13cm,angle=-90}
\\
\epsfig{file=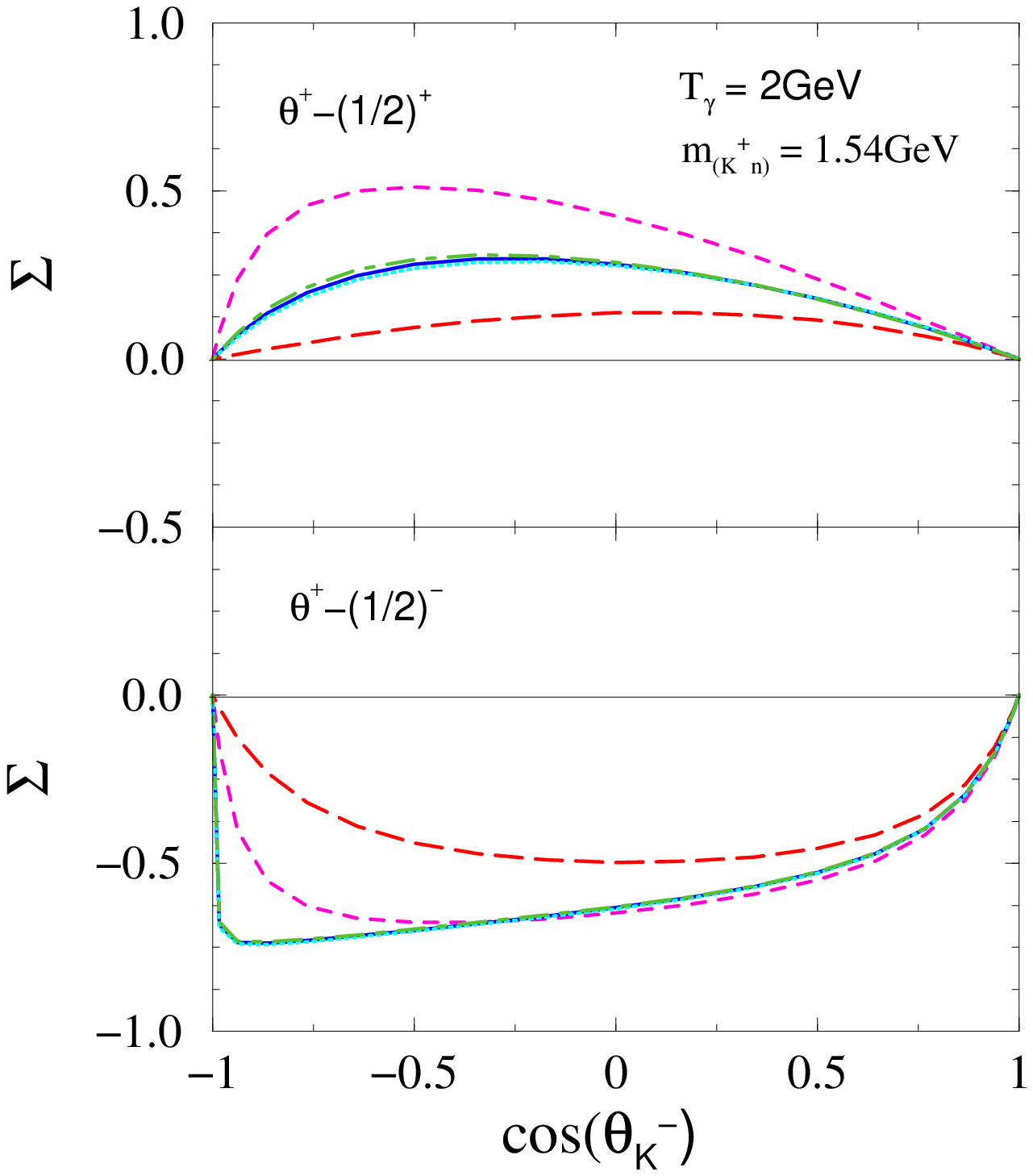,width=13cm,height=9.5cm}
\caption{The same as Fig.~\protect\ref{K_only} but with the $K+K^*$ 
contribution as defined in the text, and for some values of $g_{NK^*\Theta}$ and 
the tensor to vector coupling ratio $\kappa^*$ in Eq.(\protect\ref{NK*Theta}).
}
\label{K+Ks}
\end{center}
\end{figure}

Fig.\ref{K+Ks} illustrates the results when the $K$ and $K^*$ contributions 
($K+K^*$) are included.
Here, we show the results for fixed parameter values of
$\lambda=0$ and $\kappa_\Theta =0$ in the $K$ contribution.
Other choices of these parameters (as in Fig.\ref{K_only}) lead to 
the same qualitative conclusion and therefore we do not show the
corresponding results here. The $K^*$ contribution introduces two new 
parameters: the $NK^*\Theta$ coupling constant 
$g_{NK^*\Theta}$, and the value of the tensor to vector coupling 
constant ratio $\kappa^*$ in Eq.(\ref{NK*Theta}). We consider the 
values as given in Table~\ref{cconst}, in addition to both choices of
the sign ($\pm$) for $g_{NK^*\Theta}$. 
As one can see, both the cross section and photon asymmetry are rather 
insensitive to the values of $\kappa^*$. However, they are sensitive 
to the sign of the coupling constant $g_{NK^*\Theta}$. 
For the cross section (upper figure), 
the angular distribution changes from a strongly forward peaked
(solid curve) to a flat (dashed curve) shape as the coupling 
constant changes from $g_{NK^*\Theta}=+2.45$ to $g_{NK^*\Theta}=-2.45$.
Doubling the value of $g_{NK^*\Theta}$ (short-dashed curve) does not 
affect either the magnitude or the shape of the 
angular distribution significantly. 
For the negative parity case of $\Theta^+$ (lower panel), the effect of 
the sign of the $NK^*\Theta$ coupling constant, $g_{NK^*\Theta}=\pm 0.34$, 
is not as strong as that for the positive parity case. Also, doubling the 
coupling constant value shows no significant effect. 
The corresponding results for the photon asymmetry $\Sigma$ is shown in 
the lower figure in Fig.\ref{K+Ks}. 
For this observable the difference 
between the positive and negative parity cases is dramatic over the entire
range of the $K^-$ emission angle. The photon asymmetry $\Sigma$ 
also shows a significant sensitivity to the magnitude of 
$g_{NK^*\Theta}$. In any case, as in Fig.\ref{K_only}, the photon 
asymmetry is always positive for the positive parity choice of $\Theta^+$, 
while it is always negative for the negative parity choice. 
Thus, if $K^*$ contributes to the $\Theta^+$ excitation at all, 
it enhances the difference in the photon asymmetries between  
the positive and negative choices of the $\Theta^+$ parity.  

\begin{figure}
\begin{center}
\vspace*{-1cm}
\epsfig{file=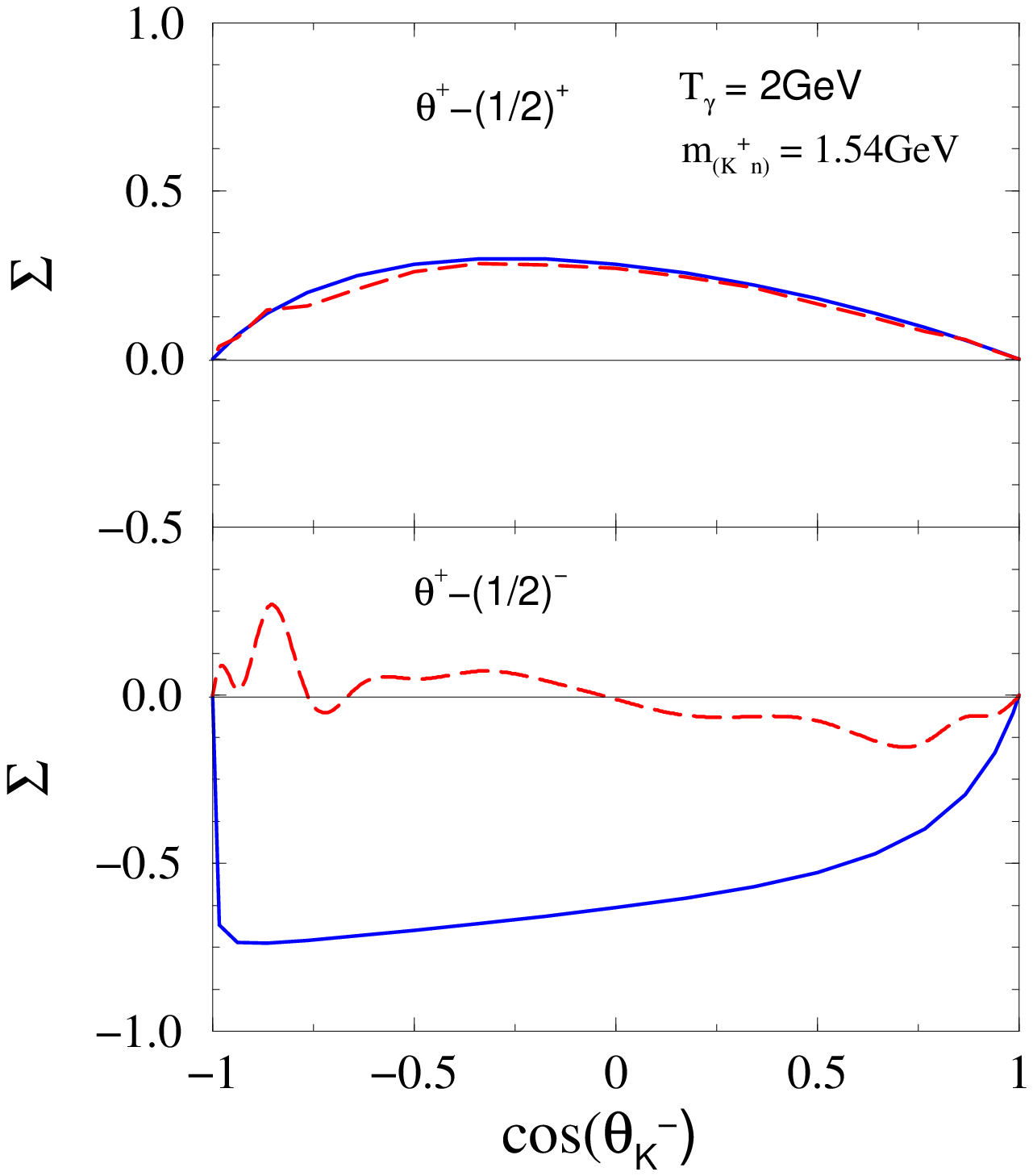,
width=13cm,height=9.5cm}
\\
\epsfig{file=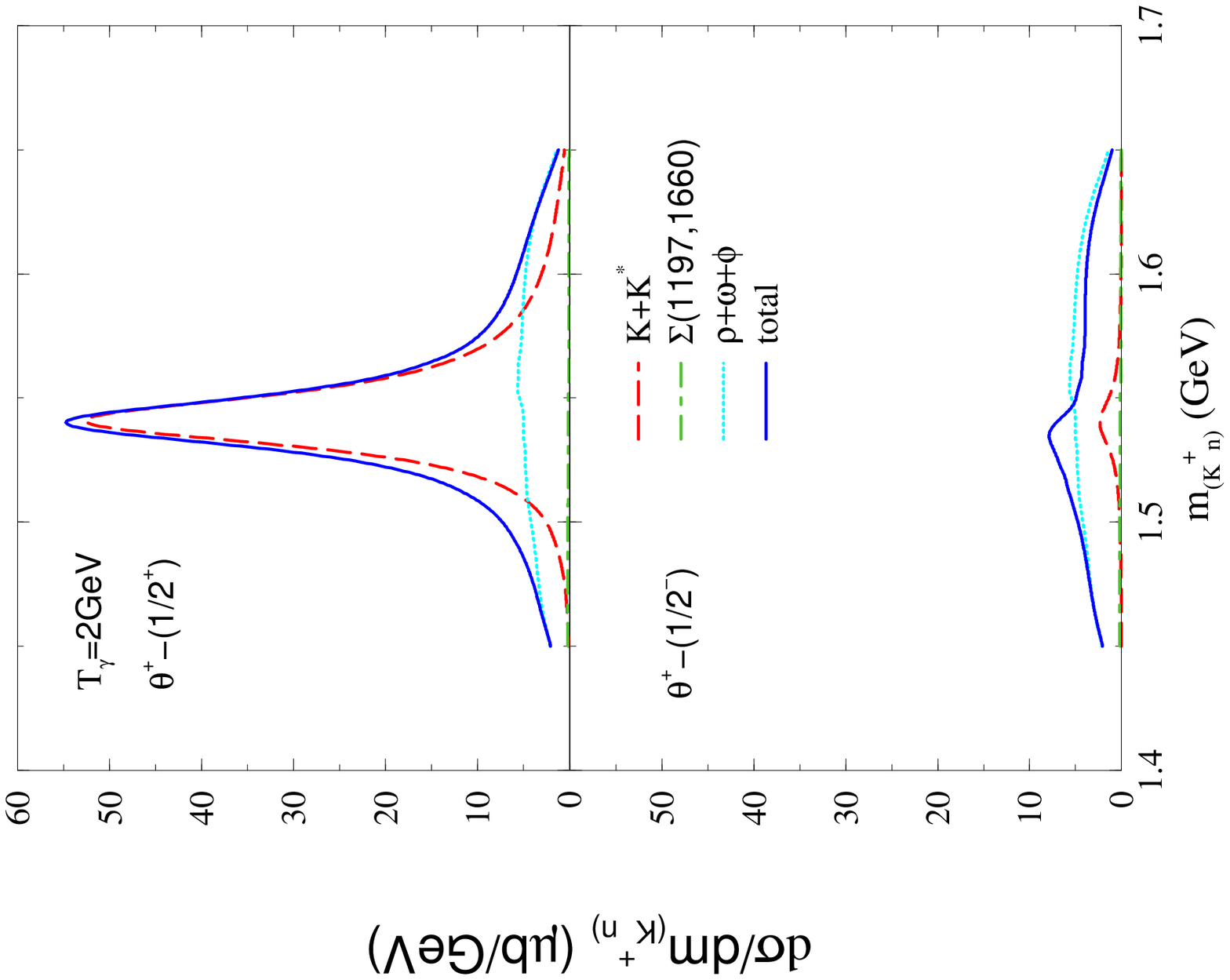,
width=10.5cm,height=13cm,angle=-90}
\caption{Upper figure: Photon asymmetry due to the $K+K^*$ contribution alone 
(solid line) and when the background is included (dashed line) due to the 
$\rho,\omega$ and $\phi$ exchanges and $\Sigma(1197)^-$ and $\Sigma(1660)^-$ 
intermediate states.
Lower figure: $(K^+ n)$ invariant mass distribution. $K+K^*$ contribution alone
(dashed curves); $\rho,\omega$ and $\phi$ exchange contribution (dotted curves);
$\Sigma(1197)^-$ and $\Sigma(1660)^-$ contribution (dash-dotted curves). The solid
curves correspond to the total contribution ($K+K^*$ + background).  
}
\label{K+Ks+background}
\end{center}
\end{figure}

We now consider the background contribution. First of all, 
as mentioned before, this contribution is largely uncertain 
theoretically, mainly due to many unknown coupling parameters
including form factors.
Some of the background contributions  
can be removed experimentally, by rejecting events associated with 
them which can be identified from appropriate
invariant/missing mass distributions~\cite{LEPS,CLAS,SAPHIR,Nakano}. 
Keeping these facts in mind, we make a rough estimate of
the background effects in the present approach. 
We consider the background consisting of the 
$\Sigma(1197)^-$ and $\Sigma(1660)^-$ intermediate states as
well as the $\rho,\omega$ and $\phi$ exchange contributions. 
The results are shown in Fig.\ref{K+Ks+background}. 
First we examine the photon asymmetry (upper figure).
The dashed curves correspond to the
results including the background contribution, while 
the solid curves correspond to the results without it.
The latter are the same results shown as solid curves 
in Fig.\ref{K+Ks}.
For the positive parity choice for $\Theta^+$, the background only
weakly affects the photon asymmetry. This is because the $K+K^*$ 
contribution is much larger than the background. 
For the negative parity case, however, the background contribution 
is relatively large and affects the photon asymmetry significantly. 
In particular, this observable becomes small in magnitude, even changing its
sign at backward angles when compared to the results without the background.  
This makes it difficult to distinguish it from the positive parity 
choice. (The wiggles shown are due 
to the pole structure of the $\phi$ meson exchange.
In the present calculation, we have taken the widths of the exchanged 
vector mesons as given in Ref.~\cite{PDG} into account.) 
The results corresponding to other choices of the parameters considered 
in the present work are not different qualitatively from those shown in 
Fig.\ref{K+Ks+background}.
Of course, how much the background will affect the resulting photon 
asymmetry depends on how large or how small it is compared to the 
$K+K^*$ ($\Theta^+$ excitation) contribution. 
Therefore, it is crucial to be able to make a reliable 
estimate of the background relative to the $K+K^*$ contribution if
we are to determine the parity of $\Theta^+$ from photon asymmetry. 
Here, we show that a measurement of the $(K^+n)$ invariant mass 
distribution can be 
used to cross check the background contribution. In the lower figure in 
Fig.~\ref{K+Ks+background} the result for the $(K^+n)$ invariant mass 
distribution is shown.        
The dashed curves correspond to the $K+K^*$ contribution alone. The dotted
curves correspond to the vector meson exchange contributions due to 
the $\rho,\omega$ and $\phi$ mesons; the dash-dotted curves correspond to the 
$\Sigma(1197)^- + \Sigma(1660)^-$ contribution. 
The solid curves denote the total contribution. As can be seen, the background 
is dominated by the exchange of vector mesons. 
For the case of positive parity $\Theta^+$ the peak-to-background ratio 
is large ($\sim 10$), and the peak structure due to $\Theta^+$  
is very clear. In contrast, for the negative parity choice this ratio is 
relatively small ($\sim 1.4$) and the peak structure due to $\Theta^+$ is 
not pronounced as for the positive parity case. This is due to 
the fact that the $K+K^*$ contribution to the cross section is
suppressed to a large extent in the case of a negative parity $\Theta^+$ 
compared to a positive parity $\Theta^+$. 
This result illustrates that,  
even if the measured photon asymmetry is small in magnitude 
(which means a significant background contribution), 
one can still learn about the parity of $\Theta^+$ 
by cross checking the peak-to-background ratio in 
the $(K^+ n)$ invariant mass distribution.
Thus, a combined analysis of the $(K^+ n)$ invariant mass distribution 
and the photon asymmetry should be able to fix the parity of $\Theta^+$. 
(Of course, measurement of the signal/noise ratio depends sensitively 
on the width of the $\Theta^+$ baryon.)   
                                                                 
In summary, we have demonstrated that measurements of photon asymmetry 
as a function of $K^-$ emission angle 
in the $\gamma n \to K^- K^+ n$ reaction, can most likely determine 
the parity of the newly discovered $\Theta^+$ pentaquark. We predict that 
if the parity of $\Theta^+$ is positive, the photon asymmetry is 
significantly positive; if the parity is negative, the photon asymmetry is 
significantly negative. It is possible that the photon asymmetry 
can be affected considerably if the background contribution is relatively 
large. Unfortunately, at present, the background contribution is not well
understood theoretically. 
In particular, for the negative parity case, the photon asymmetry 
may become very small in magnitude, even changing its sign depending on the 
$K^-$ emission angle. However, even in this worst case scenario, 
a combined analysis of the $(K^+ n)$ invariant mass distribution 
and photon asymmetry should allow a determination of the parity 
of $\Theta^+$.

After the completion of this work, a preprint by Q. Zhao and J. S. 
Al-Khalili [hep-ph/0310350] came to our attention which explores 
the sensitivity of the photon asymmetry to the parity of $\Theta^+$ in the 
$\gamma n \to K^- \Theta^+$ reaction. Their finding is consistent with
the present results.

\vspace{2em}
\noindent
{\bf Acknowledgment:}\\
\noindent
We would like to thank J. Haidenbauer for helpful discussions concerning 
the $K-N$ scattering models, and T. Nakano for information about 
experimental aspects of the $\gamma n \to K^- K^+ n$ reaction. 
We also thank W.G. Love for a careful reading of the manuscript. 
This work is supported by Forschungszentrum-J\"{u}lich, 
contract No. 41445282 (COSY-058).



%
\newpage
\begin{table}[hbtp]
\begin{center}
\caption{SU(3) relations among the coupling constants, values and types 
for the coupling constants used in the calculation. 
}
\begin{tabular}{ll}
Coupling constant &Sources\\
\hline\hline
$g_{NK\Theta} = 4.9,\hs\lambda=(0,1)=$(pv,ps) 
&$J^p = (1/2)^+, \Gamma_{(\Theta^+ \to N K)} = 25$ MeV  
Ref.~\protect\cite{LEPS}
\\
$g_{NK\Theta} = 0.7,\hs\lambda=(0,1)=$(vector,scalar)
&$J^p = (1/2)^-, \Gamma_{(\Theta^+ \to N K)} = 25$ MeV 
Ref.~\protect\cite{LEPS} 
\\
$g_{NK^*\Theta} = -g_{NK\Theta}, (1/2) g_{NK\Theta},
\hs\kappa^* = -3,0,3$
&SU(3)/empirical  $g_{NK^*Y}/g_{NKY}$ ($Y=\Sigma,\Lambda$)
\\
$g_{NK\Sigma(1197)} = -g_{\pi NN}(1 - 2\beta) = 3.6$, $\lambda=0$ 
&SU(3) + OZI ($\beta \simeq 0.63$) ~\protect\cite{Nak2,Rat} 
\\
$g_{NK^*\Sigma(1197)} = -g_{\rho NN}(1 - 2\beta) = -3.36$
&SU(3) + OZI $(\beta=0)$ ~\protect\cite{Nak1}
\\
$\kappa_{NK^*\Sigma(1197)} = 0$
&Assumption
\\
$g_{NK\Sigma(1660)} = 2.6$ 
&$g_{NK\Sigma(1660)}>0$ assumption, $\Gamma_{(\Sigma(1660) \to N\bar{K})}$ 
PDG~\protect\cite{PDG}
\\
$\kappa_{NK^*\Sigma(1660)} = 0$ &Assumption
\\
$g_{\rho NN} = 3.36,$
\hs$\kappa_\rho = 6.1$
&Bonn Potential~\protect\cite{Bonn}
\\
$g_{\omega NN} = 9.0,$
\hs$\kappa_\omega = 0$
&SU(3)+OZI~\protect\cite{Nak1,NNvec} 
\\
$g_{\phi NN} = -0.65,
\hs\kappa_\phi = 0$
&SU(3)+OZI~\protect\cite{Nak1,NNvec}
\\
$g_{\phi KK} = \frac{1}{2}(\sqrt{2}\cos\alpha_V - \sin\alpha_V)g = 4.5$
&SU(3)+OZI ~\protect\cite{Nak1}, $\Gamma_{(\phi \to K^+K^-)}$ PDG~\protect\cite{PDG}
\\
$g_{\omega KK} = \frac{1}{2}(\cos\alpha_V - \sqrt{2}\sin\alpha_V)g = 3.7$
&$\alpha_V \simeq 3.8^\circ$~\protect\cite{Nak1,Nak2}
\\
$g_{\rho KK} = \frac{1}{2}g = 3.4 $
&$g > 0$ from Vector Meson Dominance  
\\
\hline
$\kappa_{\Theta^+} = -1.8,0,1.8$ &
\\
$\kappa_{\Sigma(1197)^-} = -1.16$ &PDG~\protect\cite{PDG}
\\
$\kappa_{\Sigma(1660)^-} = \kappa_{\Sigma(1197)^-}$ 
&Assumption
\\
$g_{K^\pm K^{*\pm}\gamma} = 0.47e = 0.14$
&SU(3)+OZI ~\protect\cite{Nak1}, $\Gamma_{(K^{*\pm} \to K^\pm \gamma)}$ PDG~\protect\cite{PDG}
%
\end{tabular}
\label{cconst}
\end{center}
\end{table}
\newpage
\newpage
\newpage
\newpage
%
\end{document}